\documentclass[aps,prll,twocolumn,superscriptaddress,amsmath,amssymb]{revtex4}

\usepackage{natbib}
\usepackage{graphicx}
\usepackage{dcolumn}
\usepackage{bm}
\usepackage{hyperref}
\usepackage[mathlines]{lineno}
\usepackage[dvipsnames]{xcolor}

\begin{document}


\title{Tuning the crystalline electric field and magnetic anisotropy along the CeCuBi$_{2-x}$Sb$_{x}$ series}
	\author{G. S. Freitas}
	\affiliation{``Gleb Wataghin'' Institute of Physics, University of Campinas - UNICAMP, Campinas, S\~ao Paulo 13083-859, Brazil}

	\author{M. M. Piva}
	\affiliation{``Gleb Wataghin'' Institute of Physics, University of Campinas - UNICAMP, Campinas, S\~ao Paulo 13083-859, Brazil}
	\affiliation{Max Planck Institute for Chemical Physics of Solids,  Nöthnitzer   Str. 40, 01187 Dresden, Germany)}

	\author{R. Grossi}
	\affiliation{``Gleb Wataghin'' Institute of Physics, University of Campinas - UNICAMP, Campinas, S\~ao Paulo 13083-859, Brazil}

	\author{C. B. R. Jesus}
	\affiliation{ Departamento de Física, Universidade Federal de Sergipe - UFS, Itabaiana, Sergipe 49100-000, Brazil.}

	\author{J. C. Souza}
	\affiliation{``Gleb Wataghin'' Institute of Physics, University of Campinas - UNICAMP, Campinas, S\~ao Paulo 13083-859, Brazil}

	\author{D. S. Christovam}
	\affiliation{``Gleb Wataghin'' Institute of Physics, University of Campinas - UNICAMP, Campinas, S\~ao Paulo 13083-859, Brazil}	

	\author{N. F. Oliveira Jr.}
	\affiliation{Instituto de F\'isica, Universidade de S\~ao Paulo - USP, S\~ao Paulo-SP, 05508-090, Brazil.}

	\author{J. B. Le\~ao}
	\affiliation{NIST Center for Neutron Research, National Institute of Standards and Technology,	Gaithersburg, MD 20899-6102.}

	\author{C. Adriano}
	\affiliation{``Gleb Wataghin'' Institute of Physics, University of Campinas - UNICAMP, Campinas, S\~ao Paulo 13083-859, Brazil}
	
	\author{J. W. Lynn}
	\affiliation{NIST Center for Neutron Research, National Institute of Standards and Technology,	Gaithersburg, MD 20899-6102.}
	
	\author{P. G. Pagliuso} 
	\affiliation{``Gleb Wataghin'' Institute of Physics, University of Campinas - UNICAMP, Campinas, S\~ao Paulo 13083-859, Brazil}

\date{\today}

\begin{abstract}
We have performed X-ray powder diffraction, magnetization, electrical resistivity, heat capacity and inelastic neutron scattering (INS) to investigate the physical properties of the intermetallic series of compounds CeCuBi$_{2-x}$Sb$_{x}$. These compounds crystallize in a tetragonal structure with space group $P4/nmm$ and present antiferromagnetic transition temperatures ranging from 3.6 K to 16 K. Remarkably, the magnetization easy axis changes along the series, which is closely related to the variations of the tetragonal crystalline electric field (CEF) parameters. This evolution was analyzed using a mean field model, which included anisotropic nearest-neighbor interactions and the tetragonal CEF Hamiltonian. The CEF parameters were obtained by fitting the magnetic susceptibility data with the constraints given by the INS measurements. Finally, we discuss how this CEF evolution can affect the Kondo physics and the search for a superconducting state in this family.

\end{abstract}

\pacs{Valid PACS appear here}
\maketitle


\section{\label{sec:level1}Introduction}

The family CeTX$_{2}$ (T = transition metals; X = Bi and Sb) is an intensively studied series of tetragonal intermetallic compounds that host, in many cases, complex magnetic behavior, non-trivial crystalline electric fields (CEF) effects and competing exchange interactions, which can tune their magnetic ground states from antiferromagnetic (AFM) to ferromagnetic (FM) \cite{Adriano2014,Adriano2015,Thamizhavel2003, Muro1997, Gautreaux2009, Rosa2015, Jobiliong2005}. For instance, one can highlight the antiferromagnetic compounds with dominant CEF effects and metamagnetic transitions, such as Ce(Cu,Au)Bi$_{2}$ and CeCuSb$_{2}$ \cite{Adriano2014, Adriano2015, Thamizhavel2003, Muro1997, Gautreaux2009}; the ferromagnets CeNiSb$_{2}$ and CeCd$_{0.7}$Sb$_{2}$ \cite{Rosa2015,Thamizhavel2003}; and the peculiar compound CeAgSb$_{2}$, which presents both AFM and FM ordering at low temperatures \cite{Jobiliong2005}. Usually, a series that presents easily tunable magnetic properties from AFM to FM may provide an interesting playground to search for the emergence of interesting physical phenomena, especially in the crossover region between the two phases.  

Trying to follow this path, the antiferromagnetic CeCuBi$_{2}$ compound is a particular member of this series that orders antiferromagnetically at $T_{N}$ = 16 K with moments along the \textit{c}-axis, and has a tetragonal ZrCuSi$_{2}$-type structure ($P4/nmm$ space group). Its magnetic structure presents an antiferromagnetic coupling between the Ce$^{3+}$ ions along the \textit{c}-axis and a ferromagnetic coupling within the \textit{ab}-plane \cite{Adriano2014}. In addition, this compound possesses a spin-flop transition for a field of 5.5 T applied along the \textit{c}-axis that corresponds to the breakdown of the antiferromagnetic coupling. The large magnetic anisotropy and the weak Kondo behavior in CeCuBi$_{2}$ suggest dominant CEF effects and anisotropic RKKY (Ruderman–Kittel–Kasuya–Yosida) interactions in this compound. Furthermore, the breakdown of the de Gennes scaling observed for RECuBi$_{2}$ (RE = rare-earth)\cite{Jesus2014} usually indicates a complex and non-trivial competition between RKKY interactions and tetragonal CEF \cite{Pagliuso2001}. Recently, the ferromagnetic CeCd$_{0.7}$Sb$_{2}$ was reported with a $T_{C}$ = 3 K. Similarly to CeCuBi$_{2}$, this compound presents a large magnetic anisotropy and a weak heavy fermion behavior, suggesting that the CEF effects and RKKY interactions are also dominant for this compound \cite{Rosa2015}.

Another aspect of the CeTX$_{2}$ family is that they share some similarities with the well known families of Ce-based heavy fermion (HF) superconductors CeMIn$_{5}$ (M = Co, Rh, Ir) and Ce$_{2}$MIn$_{8}$ (M = Co, Rh, Pd), such as the crystalline structure and high magnetic anisotropy \cite{Thompson2012}. As such, chemical substitution and applied pressure may also be used to tune the Kondo effect, CEF parameters and exchange interaction in these systems toward to an unconventional superconductivity phase \cite{Piva2018}.

In this work we have combined X-ray powder diffraction (XPD), magnetization, electrical resistivity, heat capacity and inelastic neutron scattering (INS) to investigate the physical properties of the intermetallic series of compounds CeCuBi$_{2-x}$Sb$_{x}$ (\textit{x} = 0 to 2). Our measurements revealed an unusual evolution of the magnetization easy direction for the series, which is closely related to the changes in the CEF parameters. The interplay between the reported CEF evolution and its effect on the Kondo physics is discussed in comparison to other families of Ce-based tetragonal heavy fermion compounds where unconventional superconductivity has been found \cite{Thompson2012}.

\section{\label{sec:level2}Experimental details}

Single crystals of CeCuBi$_{2-x}$Sb$_{x}$ were obtained using the flux-growth method with high pure elements ($\sim$ 99.99\%) with the composition Ce:Cu:Sb:Bi = 1:1:\textit{x}:20-\textit{x} if $x \leq 1$ and 1:1:20-\textit{x}:\textit{x} otherwise. The sealed quartz tube was heated up to 1050 $^\circ$C for 12 h and then cooled down at 15 $^\circ$C/h. The excess of Bi flux was removed at 550 $^\circ$C and the Sb flux was removed at 670 $^\circ$C.  For all samples, the tetragonal crystal structure was determined by room temperature XPD and the elemental analysis was performed using both wavelength dispersive spectroscopy (WDS) and energy dispersive spectroscopy (EDS) to obtain the actual Sb concentration in the studied compounds with an error of 5\%. These are the \textit{x}'s values used along this work. The EDS measurements have shown a Cu-site vacancy of 14(3)\% in our CeCuSb$_2$ samples. In fact, the observed physical properties reported here are in good agreement with previously reported Cu deficient CeCuSb$_2$ samples \citealp{Gautreaux2009}. However, it is worth mentioning that this Cu stoichiometry variation does not invalidate and/or interfere on this work main results and conclusions about the evolution of the crystalline electric field and magnetic anisotropy along the CeCuBi$_{2-x}$Sb$_{x}$ series.

Magnetization measurements were performed using a commercial superconducting quantum interference device (SQUID) magnetometer. Specific heat measurements were done using a commercial small mass calorimeter that employs a quasiadiabatic thermal relaxation technique. The in-plane electrical resistivity was obtained using a \textit{dc} resistance bridge in a four-contact configuration. The INS experiments were performed using the BT-7 double-focusing triple-axis spectrometer \cite{Lynn2011} located at NIST Center for Neutron Research (NCNR). The masses of the samples used for the INS experiments were approximately 2 g and 6 g for the \textit{x} = 0 and \textit{x} = 0.6 compounds, respectively. Neutron scattering spectra were collected as a function of energy transfer (2 - 30 meV) and temperatures (5, 20, 50 and 100 K), with a vertically focused PG(002) monochromator and horizontally focused PG(002) analyzer fixed at 14.7 meV. In order to determine the phonon contribution to the INS data, we have performed low-Q and high-Q experimental runs. To improve statistics of the countings for the \textit{x} = 0 compound, we have taken advantage of the nondispersive nature of the CEF scattering and used non-powdered single crystals to get a larger sample size. To increase resolution in the low energy range for selected Sb doped samples, we have also performed INS experiments with a low-Q configuration at different temperatures (6, 15, 30 and 50 K) at the cold neutrons Spin-Polarized Triple-Axis Spectrometer (SPINS) at the NCNR, using a vertically focused pyrolytic-graphite (PG) monochromator to select incident neutrons with longer wavelengths.

\section{\label{sec:level3}Results and discussions}

\begin{figure}[h]
 	\includegraphics[width=0.45\textwidth]{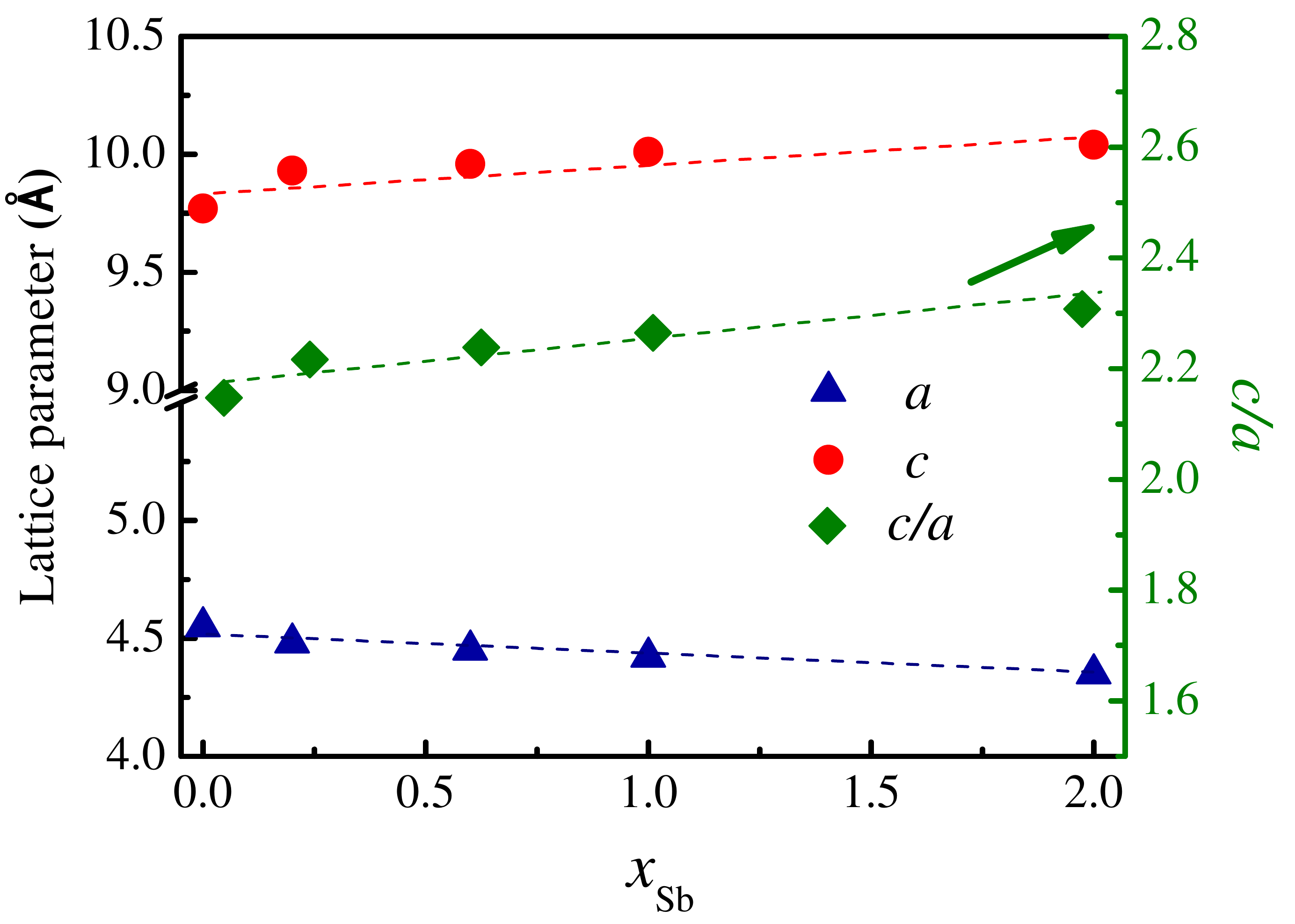}
 	\caption{\label{raiox} Tetragonal lattice parameters \textit{c} and \textit{a} (and their ratio \textit{c/a}) determined by x-ray diffraction at room temperature as a function of the Sb concentration for the series CeCuBi$_{2-x}$Sb$_{x}$ compounds.}
 \end{figure}

Figure \ref{raiox} presents the evolution of the tetragonal lattice parameters which were extracted from Rietveld refinements of the XPD data and confirm that all the samples crystallize in a ZrCuSi$_2$-type structure. We observe a reduction of the lattice parameter $a$ (consistent with the smaller size of Sb) and an increase of $c$ with respect to the \textit{x} = 0 compound. As a consequence, one can see an increase in the c/a ratio along the series.

\begin{figure*}[!t]
 	\includegraphics[width=0.8\textwidth]{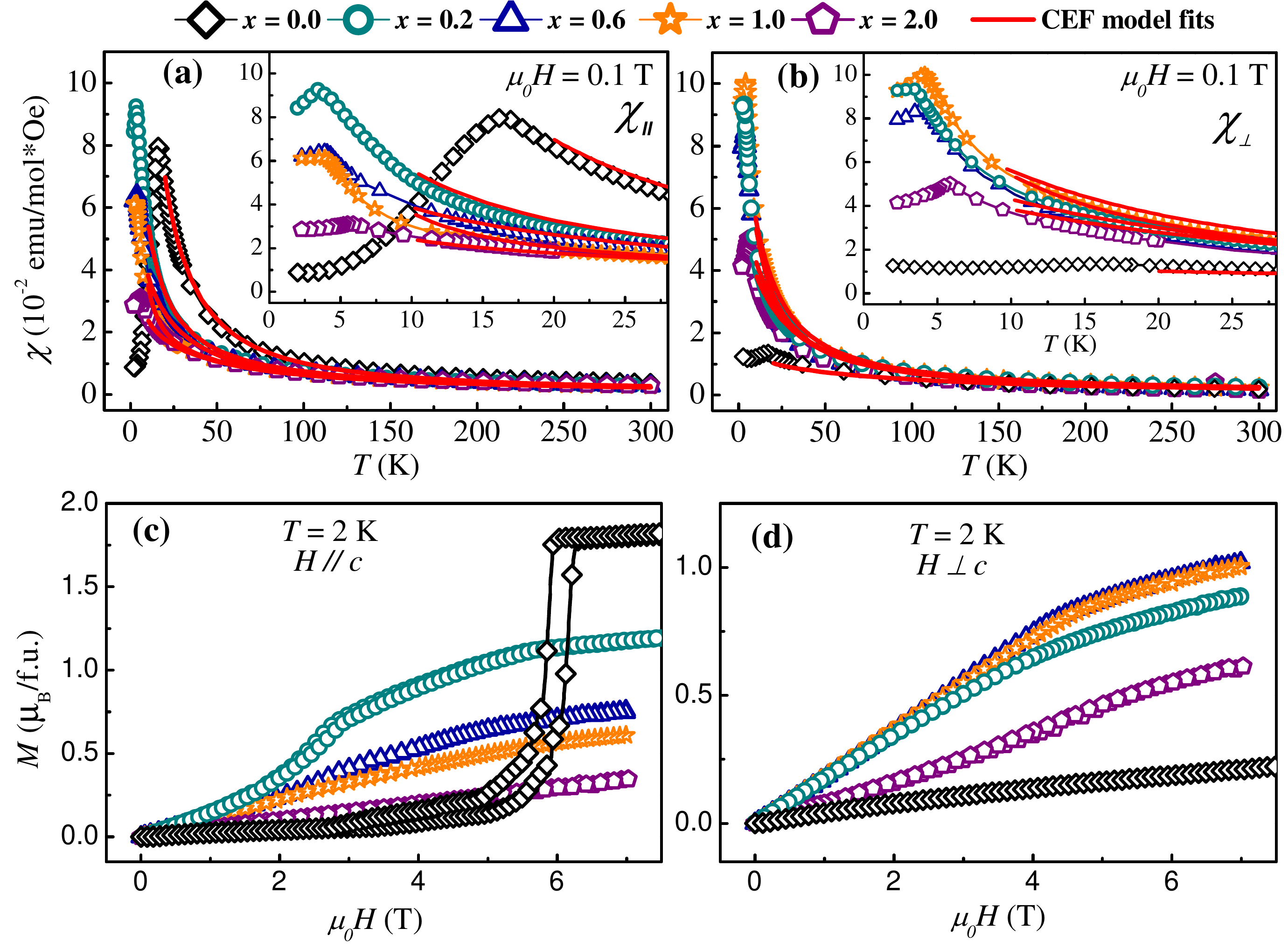}
 	
 	\caption{\label{magnetization} Temperature dependence of the magnetic susceptibility (log-linear) measured with a field of 0.1 T applied (a) parallel ($\chi_{\parallel}$) and (b) perpendicular ($\chi_{\perp}$) to the \textit{c}-axis for CeCuBi$_{2-x}$Sb$_{x}$. Field dependence of magnetization measured with the magnetic field applied (c) parallel and (d) perpendicular to the \textit{c}-axis. The curves of the compounds \textit{x} = 0 and 0.2 with the magnetic field applied parallel to the \textit{c}-axis was performed at 0.6 and 1.3 K, respectively, with the magnetic field of 0 - 17.0 T. The red solid lines in (a) and (b) are the best fits of the data using the CEF mean field model discussed in the text. [Note: 1 emu = 1 G cm$^{3}$ = 10$^{-3}$ A m$^2$].} 
 \end{figure*}

 Figure \ref{magnetization}(a)-(b) shows the temperature dependence of the magnetic susceptibity $\chi(T)$ for a magnetic field of $\mu_{0}$\textit{H} = 0.1 T applied parallel ($\chi_{\parallel}$) and perpendicular ($\chi_{\perp}$) to the \textit{c}-axis. The data reveal an AFM order for all samples, with $T_{N}$ $\simeq$ 16 K for \textit{x} = 0 and $T_{N}$ $\simeq$ 3.7 K, 3.9 K, 4.1 K and 5.6 K for the samples with \textit{x} = 0.2, 0.6, 1.0 and 2.0, respectively. One can also notice that the large magnetic anisotropy observed for CeCuBi$_{2}$, in which the magnetization easy axis is along the \textit{c}-axis, reduces as a function of Sb-concentration for CeCuBi$ _{2-x} $Sb$ _{x} $ up to \textit{x} = 0.6. At this concentration the magnetization easy axis changes to the \textit{ab}-plane, then for higher values of  \textit{x}, the magnetic anisotropy starts to increase again but with the magnetization easy axis remaining in the \textit{ab}-plane. The ratio $\chi_{\parallel}/\chi_{\perp}$ at $T_{N}$ along the series is equal to 4.5, 1.1, 0.8, 0.6 and 0.6 for \textit{x} = 0, 0.2, 0.6, 1.0 and 2.0, respectively. The T$_{N}$ and the magnetic anisotropy ratio found for the CeCu$_{1-y}$Sb$_{2}$ are consistent with the previously reported Cu-deficient compound \cite{Gautreaux2009}.    
 
 From Curie-Weiss fits to the polycrystalline average of the magnetic susceptibility data for \textit{T} $>$ 150 K, we extract an effective magnetic moment of $\mu_{eff}$ = 2.5(1) $\mu_{B}$ for all compounds in the series, consistent with the theoretical value of 2.54 $\mu_{B}$ for the Ce$^{3+}$ ion. Additionally, we show in Table \ref{valorcef} the paramagnetic Curie-Weiss temperatures ($\theta_{CW}$) obtained from these fits. The observed evolution of the magnetic anisotropy along the series, which becomes nearly isotropic for \textit{x} = 0.2, strongly indicates interesting changes of the Ce$^{3+}$ (\textit{J} = 5/2) CEF schemes. 
 
 Figure \ref{magnetization}(c)-(d) displays the low temperature magnetization as function of the applied magnetic field \textit{M(H)} for all studied compounds. The change of the magnetization easy axis in the series is also evident in these data. We can observe the suppression of the spin-flop transition observed for CeCuBi$_{2}$, when the field of 5.5 T is applied along the \textit{c}-axis ($H\parallel c$). For the doped compounds, this metamagnetic transition is rapidly suppressed. For an incorporation of 0.2 of Sb, only a small kink around 3 T can be seen and it may be associated with changes in the magnetic structure. Moreover, we see a clear increase of the magnetization for fields applied perpendicular to the c-axis ($H\perp c$) as a function of Sb substitution.

\begin{figure}[h]
 	\includegraphics[width=0.42\textwidth]{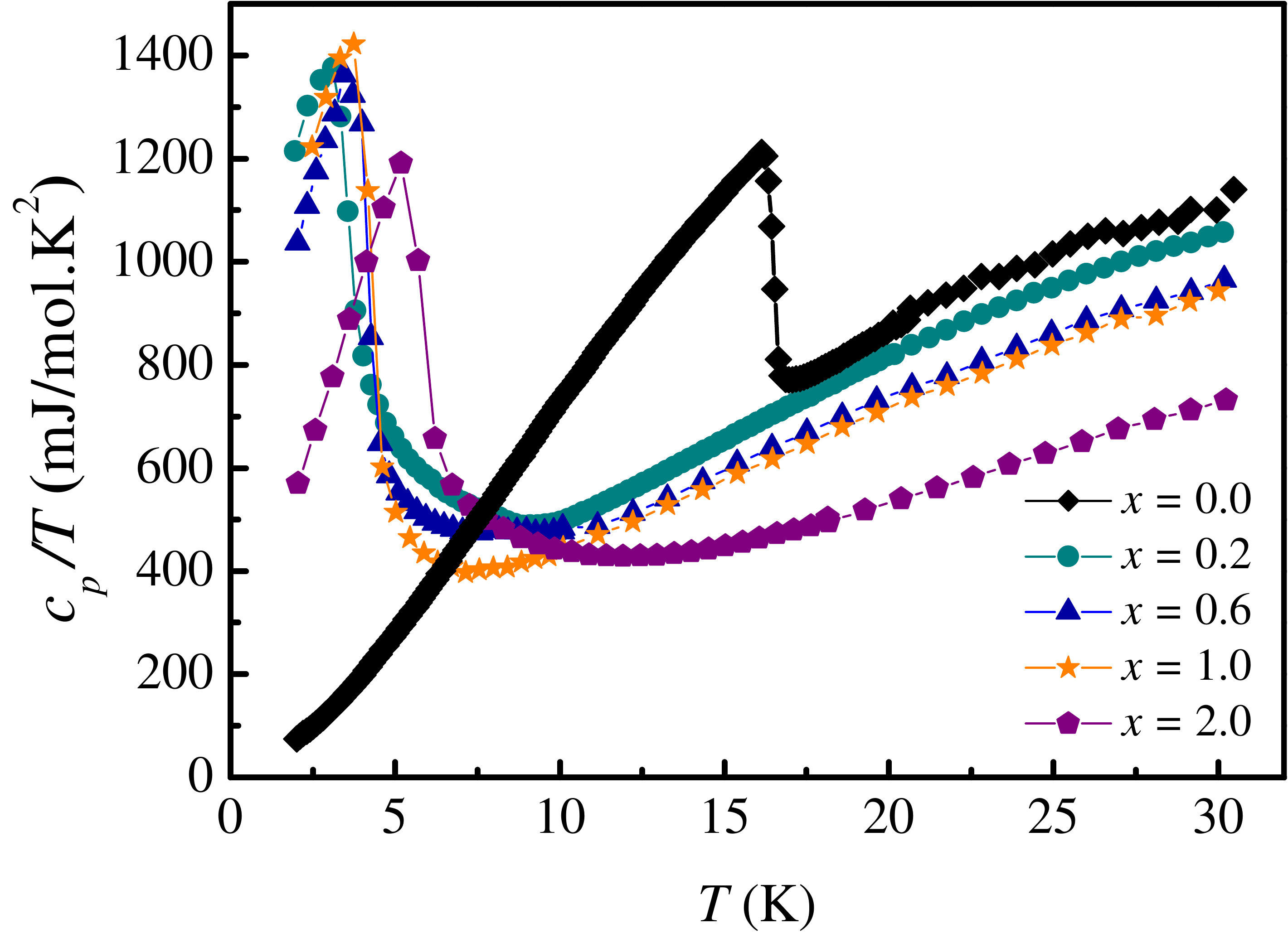}
 	\caption{\label{CxT} Specific heat as a function of the temperature for the series of CeCuBi$_{2-x}$Sb$_{x}$ compounds.}
 \end{figure}

 The specific heat divided by the temperature as a function of the temperature is displayed in Figure \ref{CxT}. The peak anomaly in the data defines $T_{N}$ = 16.5 K, 3.8 K, 4.3 K, 4.5 K and 5.6 K for \textit{x} = 0, 0.2, 0.6, 1.0 and 2.0, respectively, in excellent agreement with magnetic susceptibility merasurements (Figure \ref{magnetization}). Futhermore, the magnetic contribution of the specific heat ($c_{mag}$) can be obtained by subtracting the specific heat of the non-magnetic analog LaCuBi$_{2}$, from which we extracted the magnetic entropy ($S_{mag}$) recovered up to $T_{N}$, in units of R$\ln$(2). The obtained values for S$_{mag}$ were 80\%, 70\%, 70\%, 80\% and 70\% of R$\ln$(2) at $T_{N}$ for \textit{x} = 0, 0.2, 0.6, 1.0 and 2.0, respectively. These values suggest that the magnetic moments of the Ce$^{3+}$ CEF ground state are slightly compensated by the Kondo effect and/or by magnetic frustration effects. Additionally, we have estimated the Sommerfeld coefficient $\gamma$ using an entropy-balance construction [$S(T_{N}-\epsilon) = S(T_{N}+\epsilon)$]. The rough extracted values of $\gamma$ were $\sim$ 250 mJ/mol.K$^{2}$ for \textit{x} = 0, and then $\sim$ 1000 mJ/mol.K$^{2}$ for \textit{x} = 0.2, 0.6 and 1.0 and finally $\sim$ 700 mJ/mol K$^{2}$ for \textit{x} = 2.0. It should be noted that, since a non-magnetic analog compound is not available for the \textit{x} $>$ 0 samples to allow a proper subtraction of the phonon background contribution to the specific heat and our analyses do not include a rigorous treatment of the magnetic/Kondo excitation contributions, the $\gamma$ values reported here for the studied samples should be taken with care and only their order of magnitude and evolution should be considered.

\begin{figure}[h]
 	\includegraphics[width=0.40\textwidth]{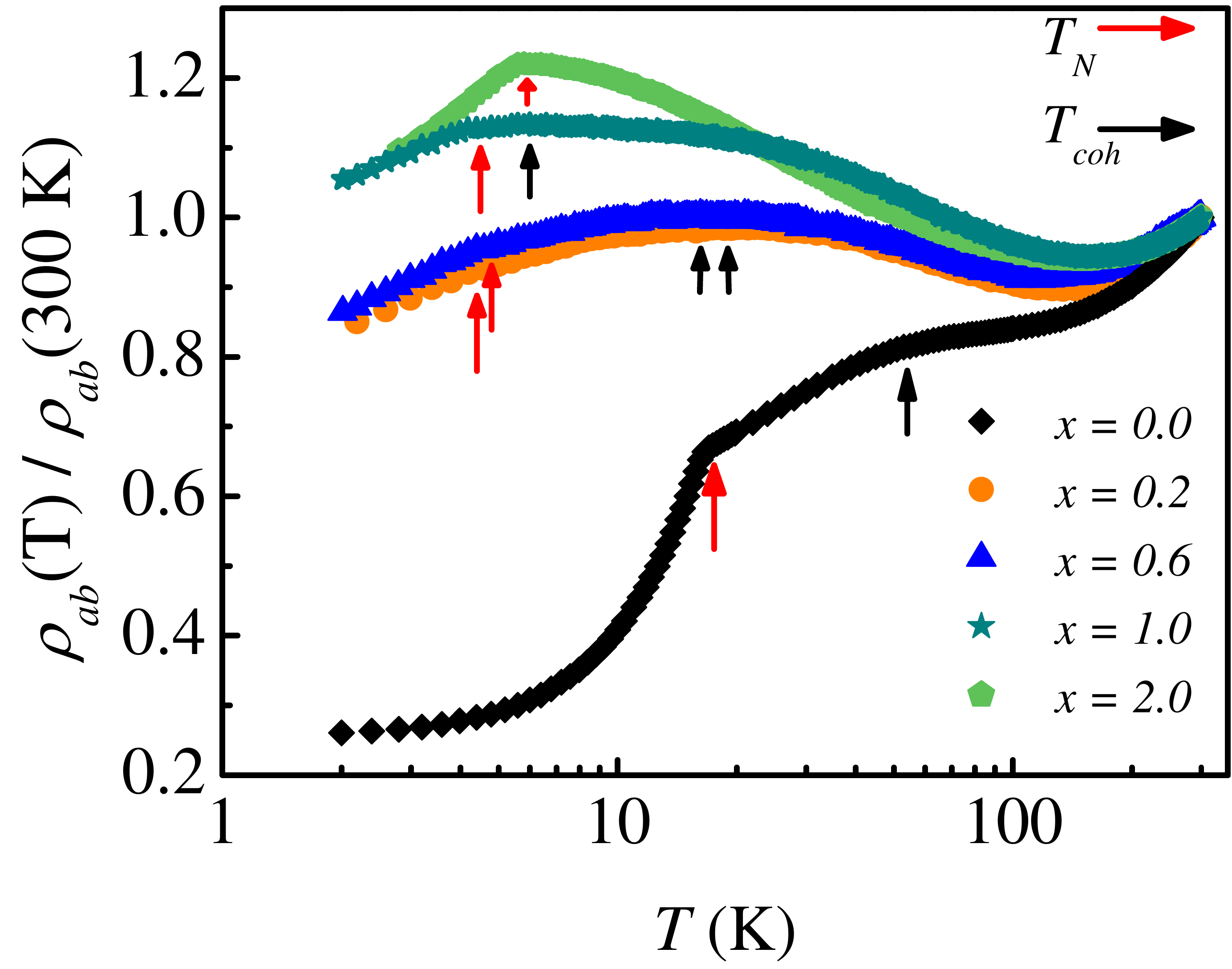}
 	\caption{\label{RxT} In-plane electrical resistivity as a function of the temperature (linear-log) for the series of CeCuBi$_{2-x}$Sb$_{x}$ compounds. The colored arrows show the evolution of T$ _{N} $ (red arrow) and T$ _{coh} $ (black arrow) throughout the series.}
 \end{figure}

  In Figure \ref{RxT} we report the in-plane electrical resistivity $\rho_{ab}$ normalized by the room temperature value as a function of temperature. The room temperature electrical resistivity values range from 80 to 160 $\mu\Omega$-cm. At lower temperatures, a small kink at $T_{N}$ is observed for each sample due to the transition to the ordered state. At higher temperatures ($T$ $>$ 150 K), $\rho_{ab}(T)$ decreases with decreasing temperature as expected for an intermetallic compound. For $T$ $<$ 150 K, we notice an increase in the resistivity, which may be related to a single impurity-like Kondo magnetic incoherent scattering \cite{Kondo}, that is typical behavior of a Kondo Lattice for temperatures above the coherence temperature. As the temperature drops, we observe the appearance of a local maximum in the resistivity, which is one of the signatures of HF compounds. The resistivity reaches this local maximum when the Kondo scattering becomes coherent at $T_{coh}$ = 50 K, 19 K, 16 K, and 6 K for x = 0, 0.2, 0.6 and 1.0, respectively. It is important to emphasize that CEF effects can also give some contribution to the formation of this maximum at $T_{coh}$ when the energy scales are comparable. For \textit{x} = 2.0 the drop in resistivity occurs only at $T_{N}$, so we could not clearly extract a $T_{coh}$ value for this sample. Interestingly, the decrease in $T_{coh}$ as a function of Sb concentration can indicate a clear evolution of the CEF effects and/or the Kondo coherence for the whole multiplet for Ce$^{3+}$ in this compounds.
 

 \begin{table*}[t]
	\caption{ N\'eel temperature, Curie-Weiss temperature, magnetic frustration parameter ($|\theta_{CW}|$/T$_{N}$), CEF parameters B$_{nm}$, isotropic exchange interaction between first (i = FN) and second neighbors, energy-level scheme and the corresponding wave functions for the series CeCuBi$_{2-x}$Sb$_{x}$.} 
	
	\centering 
	
	\begin{tabular}{c| c c c c c c c c c c c c c c c c c c c c c c c c c c c c c c c c } 
		
		\hline
		
		\textit{x} &&& {\normalsize T$_{N}$} (K) &&& {\normalsize $\theta_{CW}$} (K) &&& {\normalsize $|\theta_{CW}|$/T$_{N}$} &&& {\normalsize B$_{20}$ (K)} & & &  {\normalsize B$_{40}$ (K)} & & & {\normalsize B$_{44}$} (K) & & & {\small $z_{FN}*J_{FN}$ (K)} &&&& {\small $z_{SN}*J_{SN}$ (K) } 
		\\[0.8ex]
		\hline
		{\normalsize 0.0} &&& 16.5 &&& -23(1) &&& {\normalsize 1.4(1)} &&& {\normalsize -8.25(25)} & & & {\normalsize 0.17(2)} & & & {\normalsize 0.50(10)} & & & {\normalsize 1.34(4) } &&&& {\normalsize -1.15(10)} 
		\\[0.7ex]
		{\normalsize 0.2} &&& 3.8 &&& -13(3) &&& {\normalsize 3.4(7)} &&& {\normalsize -0.05(10)} & & & {\normalsize 0.03(1)} & & & {\normalsize 0.17(2)} & & & {\normalsize 1.41(3)} &&&& {\normalsize -0.22(5) } 
		\\[0.7ex]
		{\normalsize 0.6} &&& 4.3 &&& -5(3) &&& {\normalsize 1.2(6)} &&& {\normalsize 0.55(15)} & & & {\normalsize 0.11(1)} & & & {\normalsize -0.65(2)} & & & {\normalsize 1.85(5)} &&&& {\normalsize -0.98(7)} 
		\\[0.7ex] 
		{\normalsize 1.0} &&& 4.5 &&& -9(3) &&& {\normalsize 2.0(7)} &&&  {\normalsize 2.45(10)} & & & {\normalsize 0.085(5)} & & & {\normalsize -0.61(11)} & & & {\normalsize 1.69(5)} &&&& {\normalsize -0.66(4)} 
		\\[0.7ex] 
		{\normalsize 2.0} &&& 5.6 &&& -20(2) &&& {\normalsize 3.6(3)} &&& {\normalsize 1.85(60)} & & & {\normalsize 0.07(1)} & & & {\normalsize -0.99(9)} & & & {\normalsize 2.71(9)} &&&& {\normalsize -0.21(6)} 
		\\[0.7ex]
		
		\hline 

	\end{tabular}
	\\[0.7ex]

	\begin{tabular}{c| c c c c c c}
		\multicolumn{7}{c}{{\normalsize \textbf{Energy levels and wave functions}}}\\[0.7ex]
		
		\textit{x} && {\normalsize Ground state} && {\normalsize First excited state} && {\normalsize Second excited state}\\[0.7ex]
		\hline
		0.0 && $0.98(1)|\pm5/2\rangle - 0.21(3)|\mp3/2\rangle$ && $0.19|\pm5/2\rangle + 0.98(12)|\mp3/2\rangle$ at 63(3) K && $|\pm1/2\rangle$ at 161(4) K\\[0.7ex]
		0.2 && $0.38(5)|\pm5/2\rangle - 0.91(3)|\mp3/2\rangle$ && $0.91(3)|\pm5/2\rangle + 038(5)|\mp3/2\rangle$ at 12(1) K && $|\pm1/2\rangle$ at 13(1) K\\[0.7ex]
		0.6 && $0.40(1)|\pm5/2\rangle + 0.92(1)|\mp3/2\rangle$ && $|\pm1/2\rangle$ at 38(2) K && $0.92(1)|\pm5/2\rangle - 0.40(1)|\mp3/2\rangle$ at 50(1) K\\[0.7ex]
		1.0 && $0.28(2)|\pm5/2\rangle + 0.96(1)|\mp3/2\rangle$ && $|\pm1/2\rangle$ at 15(2) K && $0.96(1)|\pm5/2\rangle - 0.28(2)|\mp3/2\rangle$ at 60(3) K\\[0.7ex]
		2.0 && $0.43(5)|\pm5/2\rangle + 0.89(2)|\mp3/2\rangle$ && $|\pm1/2\rangle$ at 20(5) K && $0.89(2)|\pm5/2\rangle - 0.43(5)|\mp3/2\rangle$ at 64(6) K\\[0.7ex]
		\hline
	\end{tabular}
	\label{valorcef} 
\end{table*}

 In order to understand the evolution of the physical properties of the CeCuBi$_{2-x}$Sb$_{x}$ compounds, we have performed an analysis of the magnetic susceptibility data for \textit{T} $>$ 10 K using a mean-field model for the doped compounds including the anisotropic exchange interaction between nearest neighbors and the tetragonal CEF Hamiltonian with a Zeeman effect contribution:
\begin{equation}
H = B_{20}O_{2}^{0} + B_{40}O_{4}^{0} + B_{44}O_{4}^{4} + z_{i}J_{i} \cdot \langle J \rangle -  \boldsymbol{\mu}\cdot \boldsymbol{B}
\end{equation} 
 where $B_{nm}$ are CEF parameters, $O_{n}^{m}$ the Steven's operator and $ z_{i}J_{i}$ represents the exchange interactions between first (i = FN) and second neighbors (i = SN). The positive value of $ z_{i}J_{i}$ corresponds to an AFM interaction and a negative to a FM interaction. For a complete description of this theoretical model see Ref.\cite{Pagliuso2006}. In Table \ref{valorcef}, we present the CEF parameters and scheme of levels obtained from the best fits to our data. From that, one can see that Sb substitution is clearly affecting the CEF parameters in this series. For instance, one can notice a change of the $\Gamma_{7}$ CEF ground state, from a $\Gamma_{7}^{1}$: $\{\alpha|\pm5/2\rangle - \beta|\mp3/2\rangle\}$ to a  $\Gamma_{7}^{2}$: $\{\beta|\pm5/2\rangle + \alpha|\mp3/2\rangle\}$ at \textit{x} = 0.6. Also, a decrease in the contribution of the $|\pm5/2\rangle$ orbital is observed for increasing Sb concentration. Remarkably, for \textit{x} = 0.2 the CEF scheme configuration is very close to a CEF scheme of Ce$^{3+}$ (J = 5/2) in cubic symmetry, with a doublet ground state and a quartet-like state as the first excited state ($\Gamma_{7}$ and $\Gamma_{8}$) \cite{LEA1962}. This is an unexpected result for a tetragonal compound and is consistent with the small magnetic anisotropy observed for this composition in the magnetic susceptibility data (see Figure \ref{magnetization}). Additionally, for \textit{x} $\geq$ 0.6 the $\Gamma_{6}$ wave function becomes the first excited state and, for \textit{x} = 1.0, the $\Gamma_{6}$: $\{|\pm1/2\rangle\}$ is relatively close to the ground state (15 K). It is important to mention that it has been previously reported that a $\Gamma_{6}$ for the ground state favors a ferromagnetic ordering for Ce-based isostructural families of compounds \cite{Adriano2015}. For instance, both CeCd$ _{0.7} $Sb$ _{2} $ and CeAgSb$_{2}$ have a $\Gamma_{6}$ ground state \cite{Rosa2015,Jobiliong2005}. 
 
 We now discuss the effects of the CEF parameters evolution as a function of Sb substitution on the magnetic properties. Comparing the pure compounds CeCuBi$ _{2} $ and CeCuSb$ _{2} $, a similar $\theta_{CW}$ is observed, however $T_{N}$ is strongly suppressed in CeCuSb$_{2}$. Certainly, the evolution of the CEF effects as a function of Sb substitution is the key ingredient that can lead to a CEF induced magnetic frustration, which reduces $T_{N}$ for the Sb-rich compounds \cite{Pagliuso2006}. The ratio $|\theta_{CW}|$/$T_{N}$, summarized in Table \ref{valorcef}, increases as a function of Sb substitution, which reinforces the magnetic frustration scenario. Besides that, the decrease of the values $z_{i}J_{i}$ indicates a reduction of the AFM and FM interactions, opposed to the increase of the magnetic frustration, reinforcing the scenario of CEF effects dominating these frustrations.
 
\begin{figure}[h]
 	\includegraphics[width=0.5\textwidth]{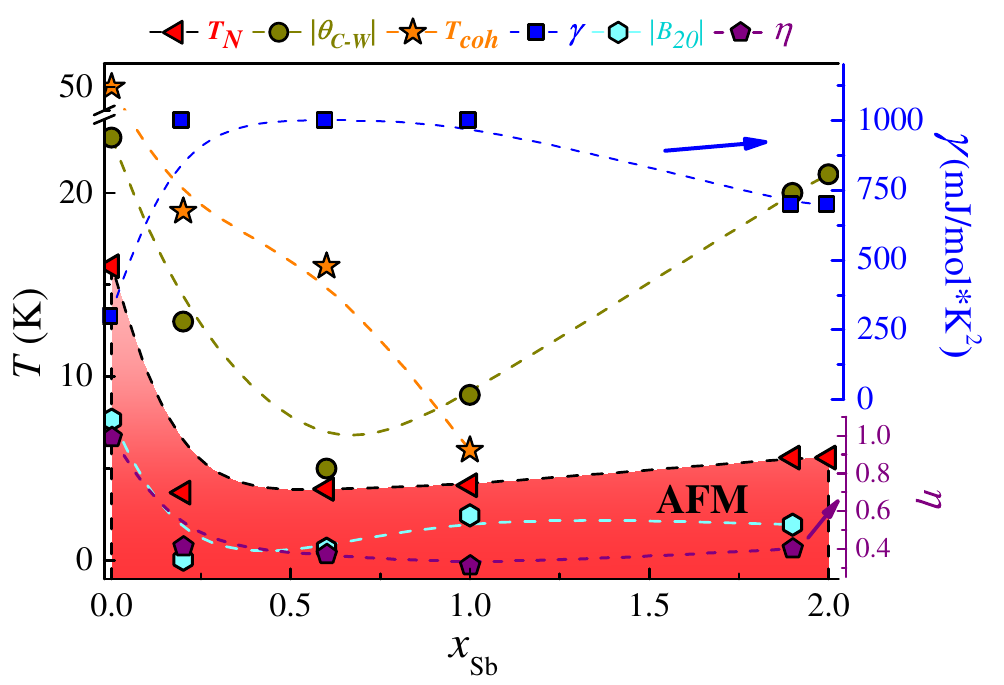}
 	\caption{Phase diagram showing the values of $T_{N}$, $|\theta_{CW}|$, $\gamma$, $T_{coh}$, $|B_{20}|$ and $\eta$ for CeCuBi$_{2-x}$Sb$_{x}$. \label{diagrama}}
 \end{figure}

 In Figure \ref{diagrama} we display a phase diagram with the most relevant physical quantities. The value of $T_{N}$ follows the behavior of $|\theta_{CW}|$ at first, however this trend is broken at \textit{x} = 1.0, primarily due to the CEF effects. The increase observed in the $\gamma$ value and the constant percentage of magnetic entropy recovered at $T_{N}$ as a function of \textit{x}, suggest that the Ce$^{3+}$ magnetic moments are being slightly compensated in the ground state by Kondo interactions, which can also contribute to the decrease of $T_{N}$. However, the suppression of $T_{coh}$ which goes into the ordered state as a function of Sb, suggests that the Kondo effect does not follow the increase of $\gamma$, which may come mainly from the CEF ground state. Interestingly, the behavior of $|B_{20}|$, which is the most relevant parameter for a tetragonal CEF hamiltonian, exhibits a remarkable trend following $T_{N}$ as a function of Sb substitution. This is in agreement with the CEF induced magnetic frustration scenario leading to the suppression of $T_{N}$ \cite{Pagliuso2006}. This can also affect $T_{coh}$ and $\gamma$ in the ground state. The $\eta$ value defined as the $|\pm5/2\rangle$ orbital contribution to the ground state, also shows an interesting evolution with $T_{N}$. Besides that, one can see that for \textit{x} $ \geq $ 0.2 the $|\pm3/2\rangle$ orbital has the most contribution in the ground state, what is not usually observed for an AFM compound of the CeTX$_{2}$ family. 

In the CeMIn$_{5}$ (M = Co, Ir and Rh) family of unconventional superconductors, compounds with a larger $|\pm3/2\rangle$ orbital contribution exhibit superconductivity at low temperatures \cite{Willers2015}. Moreover, these compounds present a magnetic anisotropy with the c-axis as the easy axis. In our case, although we have found a CEF ground state with a higher $|\pm3/2\rangle$ contribution, the CEF scheme is closer to a cubic symmetry, with a $\beta$ $\approx$ 0.9 \cite{LEA1962}. In addition, the direction of the magnetic moment of Ce$^{3+}$ ion in the series moves towards of the \textit{ab}-plane, as observed in the magnetic susceptibility measurements. Possibly, these two properties may not favor a superconducting state in CeCuBi$_{2-x}$Sb$_{x}$.

\begin{figure}[h]
	\includegraphics[width=0.48\textwidth]{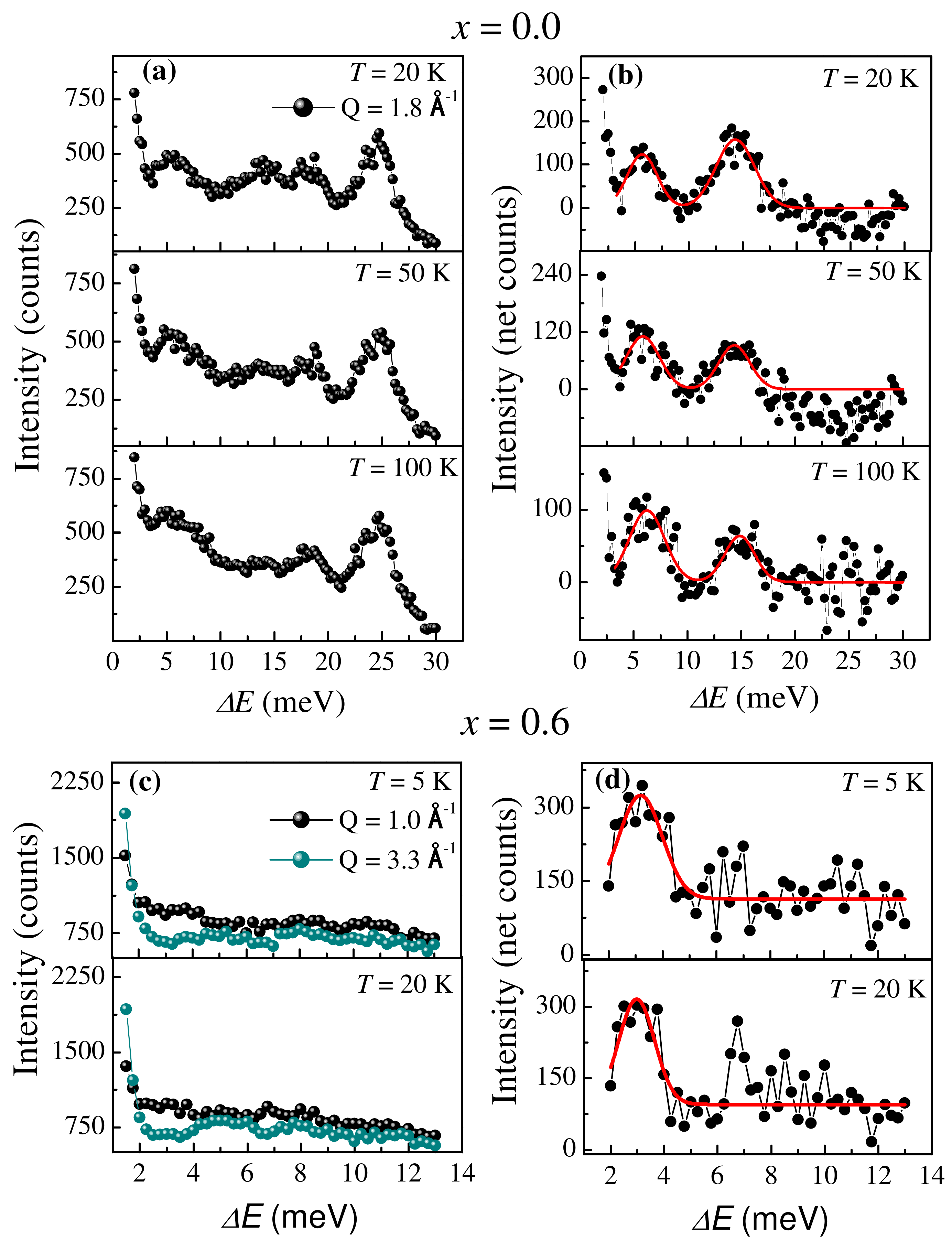}
	\caption{\label{neutrons} Inelastic neutron scattering spectra for the compound (a) CeCuBi$_{2}$ and (c) CeCuBi$_{1.4}$Sb$_{0.6}$; Subtracted data fit with Gaussians for (b) CeCuBi$_{2}$ and (d) for CeCuBi$_{1.4}$Sb$_{0.6}$ (red solid line).}
\end{figure}

To gain a microscopic insight and to confirm the obtained CEF wave functions and energy levels, we turn now to the discussion of the INS measurements performed on the \textit{x} = 0 and \textit{x} = 0.6 compounds. The data are reported in Figure \ref{neutrons} (a)-(c). For the \textit{x} = 0 compound (Figure \ref{neutrons}(a)), the INS experiments at \textit{T} = 20 K, 50 K and 100 K with a reciprocal wave vector of Q = 1.8 $ \AA^{-1} $ showed four excitation peaks at $\Delta E$ = 5.5, 14.0, 18.5 and 24.6 meV. It is important to notice, the two low energy peaks show a decrease in the intensity as the temperature increases, consistent with INS peaks that are due to CEF excitations from the ground state, whereas the high energy peaks were found to be nearly temperature independent. To confirm that the $\Delta E$ = 18.5 and 24.6 meV peaks are from phonon density-of-states scattering, measurements at higher Q (not shown) were taken. At higher Q we observed an increase in the intensity of the 18.5 and 24.6 peaks, as expected for phonon scattering, in contrast to magnetic scattering, which decreases for higher Q according to the magnetic form factor. The broadened peaks of the CEF excitations are probably due to the Kondo scattering. To further analyze the data, we subtracted from the experimental data an appropriate background of the elastic and phonon contributions to the INS spectra using a combined polynomial and Gaussian best fits to the data at 100 K. The subtracted data are shown in Figure \ref{neutrons}(b) with two Gaussian fits of the CEF excitations on  $\Delta E$ = 5.5 and 14.0 meV. 

In Figure \ref{neutrons}(c) we report the INS spectra for a low-Q (1.4 $\AA^{-1}$) and a high-Q (3.3 $\AA^{-1}$) configuration at 5 K and 20 K for the \textit{x} = 0.6 compound. We can observe a clear difference of intensity between the low-Q and high-Q spectra in the region of 2 to 4 meV (20 to 50 K, where 1 meV corresponds to 11.6 K in thermal energy.), which is probably related to CEF excitations. That excitation would be consistent with the CEF scheme proposed from the magnetic susceptibility fit, which provides two doublets at 40 K and 50 K (Table \ref{valorcef}). These values of the energy splittings together with the broad linewidths would make it difficult to distinguish more than one peak in the INS spectra, as one can observe in Figure \ref{neutrons}(c). Furthermore, as mentioned before, the Kondo effect usually causes a broadening in CEF excitation peaks, preventing a clear distinction of the possible two peaks in that small $\Delta E$ region. Therefore, the region between 2 and 4 meV could have more than one peak of CEF excitations. For this reason, we have tried to perform the INS experiment on the SPINS beamline, which has better energy resolution than the BT-7 beamline used for the experiment displayed in Figure \ref{neutrons}. However, the experiment was not conclusive due to the low neutron flux of the beamline and to the $\Delta E$ closest to a high-intensity elastic peak (data not shown).

“In the region between 6 to 8 meV, we can observe a sharp peak-like feature in the difference of the low-Q and high-Q spectra, which appears to increase as a function of temperature. If this feature were associated with CEF effects, this increase would suggest that this supposed CEF excitation would be related to a transition from a first excited state to a second excited state, giving us an energy splitting $\Delta E$ $\approx$ 7 meV. Such a CEF scheme gives rise to a larger magnetic anisotropy than the one that we have obtained in our experiments for the \textit{x} = 0.6 compound. Besides that, the peak profile after the spectra subtraction provides width values close to the instrumental resolution, leading to us to disregard this peak as a CEF excitation. Therefore, we could not fit the complete set of data with this alternative CEF scheme. For this reason, we have only considered the first peak as a CEF excitation. In Figure \ref{neutrons}(d) we showed the subtraction of the low-Q by the high-Q spectra fit using a Gaussian peak.

The Gaussian curves shown in Figure \ref{neutrons}(b) were used to compare the CEF excitations of the INS experiments with the CEF scheme obtained from the magnetic susceptibility fits. These analyses were made by correlating the ratio of each neutron scattering cross-section due to a CEF excitation, with the ratio of the integrated intensity of each CEF peak of the INS spectra. The differential cross-section for the neutron scattering due to a CEF transition from an initial to a final state used is expressed as:
 \begin{eqnarray}
\dfrac{d^{2}\sigma(i \rightarrow j)}{d\Omega dE'} = N \dfrac{k_{f}}{k_{i}} \bigg(\dfrac{\hbar\gamma e^{2}}{m c^{2}}\bigg) e^{-2W} \bigg|\dfrac{1}{2}g_{J}f(\textbf{Q})\bigg|^{2} \\\nonumber
\times\sum_{i,j}n_{i} |\langle i |J_{\perp} | j \rangle|^{2} \delta(E_{i}-E_{j}+\hbar\omega) \label{eqneutron}
\end{eqnarray}
where $f(\textbf{Q})$ is the magnetic form factor, $ k_{f} $ and $k_{i}$ are the initial and final neutron wave vectors, and $n_{i}$ is the population of the initial state. For that analysis, we have used a CEF level scheme similar to the one obtained from the magnetization fits (Table \ref{valorcef}), which is a $\Gamma_{7}^{1} \rightarrow \Gamma_{7}^{2} \rightarrow \Gamma_{6}$ in the ground state, first and second excited state, respectively. Thereby, we were able to estimate the mixing parameters $\alpha$ and $\beta$ for the \textit{x} = 0 compound, and the results are shown below. This CEF level scheme obtained by the INS experiment was used to fit the magnetic susceptibility shown in Figure \ref{magnetization} providing a small modification of the CEF parameters from the previously published CEF scheme \cite{Adriano2014}, and these updated values are presented in Table \ref{valorcef}. Since we were not able to distinguish individual peaks in the INS spectra of the \textit{x} = 0.6 compound, this analysis for the integrated intensity was not possible. The wave functions extracted in the INS data for the \textit{x} = 0 compound is shown below. The saturation magnetic moment of the proposed ground state is 1.7 $\mu_{B}$, similar to the one obtained in Figure \ref{magnetization}(c), indicating small Kondo compensation for this compound.

  \begin{gather}
	\Gamma_{7}^{1}: \{0.9(1)|\pm5/2\rangle - 0.2(2)|\mp3/2\rangle\} \text{ at 0 meV (0 K)}\nonumber\\
	\Gamma_{7}^{2}: \{0.2(2)|\pm5/2\rangle + 0.9(1)|\mp3/2\rangle\  \text{ at 5.5 meV (65 K)}\nonumber\\
	\Gamma_{6}: \{|\pm1/2\rangle\} \text{ at 14 meV (162 K)}\nonumber
\end{gather}

\section{\label{sec:level4}Conclusions}

We have presented a detailed study of the evolution of CEF effects inducing magnetic anisotropy on the series of antiferromagnetic CeCuBi$_{2-x}$Sb$_{x}$ compounds, using room temperature X-ray powder diffraction, low temperature field-dependent magnetization, magnetic susceptibility, specific heat capacity, electrical resistivity, and inelastic neutron scattering. The substitution of Bi by Sb in the compound has induced modifications of CEF parameters, which lead to a change of the magnetic easy-axis from the \textit{c}-axis to the \textit{ab}-plane. Such change is associated with an increase of the magnetic frustration, along with a consequent suppression of $T_{N}$, as a function of Sb concentration. Using a model of a tetragonal CEF Hamiltonian to fit the magnetic susceptibility measurements, we have found the evolution of Ce$^{3+}$ CEF wave functions and energy levels, which reproduces the behavior of the magnetic anisotropy found as a function of Sb content. INS experiments were performed on the compounds CeCuBi$_{2}$ and CeCuBi$_{1.4}$Sb$_{0.6}$ to successfully verify these CEF schemes. Further microscopic experiments, such as X-ray absorption and nuclear magnetic resonance, would be highly desirable to confirm the trend along the series.

\begin{acknowledgments}
This work was supported by FAPESP (in particular grants No 2015/16191-5, 2015/15665-3, 2017/10581-1, 2017/25269-3, 2018/11364-7 and 2019/04196-3), CAPES, CNPq and FINEP-Brazil. The authors acknowledge the Brazilian Nanotechnology National Laboratory (LNNano) and the Center for Semiconducting Components and Nanotechnologies (CCSNano-Unicamp) for providing the equipment and technical support for the EDS experiments.
\end{acknowledgments}

\end{document}